\documentclass[nonblindrev]{informs3} 

\usepackage{balance}
\usepackage{amsmath}
\usepackage[tight,footnotesize]{subfigure}
\usepackage{graphicx}
\usepackage{bm}
\usepackage{algorithm}
\usepackage[noend]{algorithmic}
\usepackage{multirow}
\usepackage{slashbox}
\usepackage{caption}
\usepackage{amsfonts}

\OneAndAHalfSpacedXII 

\usepackage{natbib}
 \bibpunct[, ]{(}{)}{,}{a}{}{,}%
 \def\bibfont{\small}%
\TheoremsNumberedThrough     

\EquationsNumberedThrough    

\MANUSCRIPTNO{2020} 

\begin{document}


\RUNAUTHOR{Tang et al.}

\RUNTITLE{Beyond Pointwise Submodularity:   Non-Monotone Adaptive Submodular Maximization subject to Knapsack and $k$-System Constraints}

\TITLE{Beyond Pointwise Submodularity:    Non-Monotone Adaptive Submodular Maximization subject to Knapsack and $k$-System Constraints}

\ARTICLEAUTHORS{%
\AUTHOR{Shaojie Tang}
\AFF{Naveen Jindal School of Management, The University of Texas at Dallas}
} 

\ABSTRACT{In this paper, we study the non-monotone adaptive submodular maximization problem subject to a knapsack and a $k$-system constraints. The input of our problem is a set of items, where each item has a particular state drawn from a known prior distribution. However, the state of an item is initially unknown, one must select an item in order to reveal the state of that item. There is a utility function which is defined over items and states. Our objective is to sequentially select a group of items to maximize the expected utility. Although the cardinality-constrained non-monotone adaptive submodular maximization has been well studied in the literature, whether there exists an approximation solution for the knapsack-constrained  or $k$-system constrained adaptive submodular maximization problem remains an open problem. It fact, it has only been settled given the additional assumption of pointwise submodularity. In this paper, we remove the common assumption on pointwise submodularity and propose the first approximation solutions for these problems. Inspired by two recent studies on non-monotone adaptive submodular maximization, we develop a sampling-based randomized algorithm that achieves a  $\frac{1}{10}$ approximation for the case of a knapsack constraint and that achieves a $\frac{1}{2k+4}$ approximation ratio for the case of a $k$-system constraint.
}


\maketitle

\section{Introduction}

In \cite{golovin2011adaptive}, they extend the study of submodular maximization from the non-adaptive setting \cite{nemhauser1978analysis} to the adaptive setting. They introduce the notions of adaptive monotonicity and submodularity, and show that a simple adaptive greedy policy achieves a $1-1/e$ approximation ratio if the utility function is adaptive submodular and adaptive monotone. Although there have been numerous research studies on adaptive submodular maximization under different settings \cite{chen2013near,tang2020influence,tang2020price,yuan2017adaptive,fujii2019beyond}, most of them assume adaptive monotonicity.  For the case of maximizing a non-monotone adaptive submodular function subject to a cardinality constraint, \cite{tang2021beyond} develops the first constant approximation solution. For the case of maximizing a non-monotone adaptive submodular and pointwise submodular function, \cite{amanatidis2020fast,hanicml} develop effective solutions  for the case of a knapsack and a $k$-system constraints, respectively.  Note that adaptive submodularity does not  imply pointwise submodularity and vice versa \cite{guillory2010interactive,golovin2011adaptive}, and this raises the following question: Does there exist an  approximation solution for maximizing a knapsack-constrained or a $k$-system constrained non-monotone adaptive submodular function without resorting to pointwise submodularity?

In this paper, we answer the above question affirmatively by proposing the first approximation solutions for both knapsack or a $k$-system constraints. Note that many practical constraints, including cardinality, matroid, intersection of $k$ matroids, $k$-matchoid and
$k$-extendible constraints, all belong to the family of $k$-system constraints.
In particular, we develop a $\frac{1}{10}$ approximate solution for maximizing a knapsack-constrained non-monotone adaptive submodular function. Technically speaking, our design is an extension of the classic modified density greedy algorithm \cite{wolsey1982maximising,amanatidis2020fast}. In particular, their design is required to  maintain two candidate policies, i.e., one is to choose a best singleton and the other one is to choose items in a density-greedy manner, while our design maintains three candidate policies in order to drop the common assumption about pointwise submodularity. For the case of a $k$-system constraint, we are inspired by the sampling based policy proposed in \cite{hanicml} and develop a similar policy that achieves a $\frac{1}{2k+4}$ approximation ratio without resorting to pointwise submodularity. We list the performance bounds of the closely related studies in Table \ref{rrr}.

\begin{table}[t]
\begin{center}
\begin{tabular}{ |c|c|c|c| }
\hline
Source & Ratio & Constraint & Require pointwise submodularity? \\
\hline
\cite{gotovos2015non} & $\frac{1}{e}$ &cardinality constraint & Yes \\
\cite{amanatidis2020fast}& $\frac{1}{9}$ & knapsack constraint & Yes \\
\cite{hanicml}& $\frac{1}{k+2\sqrt{k+1}+2}$ & $k$-system constraint & Yes \\
\cite{tang2021beyond} & $\frac{1}{e}$ &cardinality constraint & No \\
this work & $\frac{1}{10}$ & knapsack constraint & No \\
this work & $\frac{1}{2k+4}$ &  $k$-system constraint & No \\
\hline
\end{tabular}
\caption{Approximation for non-monotone adaptive submodular function maximization}
\label{rrr}
\end{center}
\end{table}

\section{Preliminaries}
We first introduce some important notations. In the rest of this paper, we use  $[m]$ to denote the set $\{0, 1, 2, \cdots, m\}$, and we use $|S|$ to denote the cardinality of a set $S$.
\subsection{Items and  States} We consider a set  $E$ of $n$ items, where each item $e\in E$ is in a particular state from $O$.  We use  $\phi: E\rightarrow O$ to denote a \emph{realization}, where $\phi(e)$ represents the state of $e\in E$.   Let $\Phi=\{\Phi(e) \mid e\in E\}$ denote a random realization, where $\Phi(e) \in O$ is a random realization of the state of $e\in E$.   The state of each item is unknown initially, one must pick an item $e\in E$ before observing the value of $\Phi(e)$. We assume there is a known prior probability distribution $p(\phi)=\{\Pr[\Phi=\phi]: \phi\in U\}$ over realizations $U$.  For any subset of items $S\subseteq E$, we use $\psi: S\rightarrow O$ to denote a \emph{partial realization} and $\mathrm{dom}(\psi)=S$ is called the \emph{domain} of $\psi$. Consider any realization $\phi$ and any  partial realization $\psi$, we say that $\psi$ is consistent with $\phi$, i.e., $\psi \prec \phi$, if they are equal everywhere in the domain of $\psi$. We say that $\psi$  is a \emph{subrealization} of  $\psi'$, i.e.,   $\psi \subseteq \psi'$, if $\mathrm{dom}(\psi) \subseteq \mathrm{dom}(\psi')$ and they are equal everywhere in $\mathrm{dom}(\psi)$. Moreover, we use $p(\phi\mid \psi)$ to denote the conditional distribution over realizations conditioned on  a partial realization $\psi$: $p(\phi\mid \psi) =\Pr[\Phi=\phi\mid \psi\prec\Phi ]$. There is a non-negative utility function $f$ that is defined over items and their states: $f: 2^{E}\times O^E\rightarrow \mathbb{R}_{\geq0}$.

\subsection{Policies and Problem Formulation} A typical adaptive policy works as follows: select the first item and observe its state, then continue to select the next item based on the observations collected so far, and so on. After each selection, we observe some \emph{partial realization} $\psi$ of the states of  some subset of $E$, for example, we are able to observe the partial realization of the states of those items which have been selected. Formally, any adaptive policy can be represented as a function $\pi$ that maps a set of observations  to a distribution $\mathcal{P}(E)$ of $E$: $\pi: 2^{E}\times O^E  \rightarrow \mathcal{P}(E)$, specifying which item to pick next based on the current observation.

\begin{definition}[Policy  Concatenation]
Given two policies $\pi$ and $\pi'$,  let $\pi @\pi'$ denote a policy that runs $\pi$ first, and then runs $\pi'$, ignoring the observation obtained from running $\pi$.
\end{definition}

Let the random variable $E(\pi, \phi)$ denote the subset of items selected by $\pi$ under a realization $\phi$. The expected  utility $f_{avg}(\pi)$ of a policy $\pi$ is
\begin{eqnarray}
f_{avg}(\pi)=\mathbb{E}_{\Phi\sim p(\phi), \Pi}f(E(\pi, \Phi), \Phi)
\end{eqnarray}
where the expectation is taken over $\Phi$ with respect to $p(\phi)$ and the random output of $\pi$. For ease of presentation, let $f(e)=\mathbb{E}_{\Phi\sim p(\phi)}f(\{e\}, \Phi)$.

\begin{definition}[Independence System] Given a ground set $E$ and a collection of sets $\mathcal{I}\subseteq 2^E$, the pair $(E, \mathcal{I})$ is an \emph{independence system} if
\begin{enumerate}
\item $\emptyset \in \mathcal{I}$;
\item $\mathcal{I}$, which is called the \emph{independent sets}, is downward-closed, that is, $A \in \mathcal{I}$ and $B \subseteq A$ implies that $B \in \mathcal{I}$. 
\end{enumerate}
\end{definition}

A set $B \in \mathcal{I}$ is called  a \emph{base} if $A\in \mathcal{I}$ and $B \subseteq A$ imply that $B = A$. A set $B \in \mathcal{I}$ is called a base of $R$ if  $B\subseteq R$ and  $B$ is a base of the independence system $(R, 2^R \cap \mathcal{I})$.

\begin{definition}[$k$-System] An independence system $(E, \mathcal{I})$  is a $k$-system for an integer $k\geq 1$ if for every set
$R\subseteq E$, the ratio between the sizes of the largest and smallest bases of $R$ is upper bounded by $k$.
\end{definition}

Let $\Omega$ denote the set of feasible policies and let $U^+=\{\phi\in U\mid p(\phi)>0\}$. For the case of knapsack constraint, define  $\Omega=\{\pi|\forall \phi\in U^+, \sum_{e\in E(\pi, \phi)} c_e \leq b\}$ where  $c_e$ is the cost of $e$, which is fixed and pre-known, and  $b$ is the budget constraint. For the case of $k$-system constraint, define  $\Omega=\{\pi|\forall \phi\in U^+,  E(\pi, \phi)\in \mathcal{I}\}$ where $(E, \mathcal{I})$ is a $k$-system. Our goal is to find a feasible policy  $\pi^{opt}$ that maximizes the expected utility, i.e., $\pi^{opt} \in \arg\max_{\pi \in \Omega} f_{avg}(\pi)$.

\subsection{Adaptive Submodularity and Pointwise Submodularity}
We start by introducing the conditional expected marginal utility of an item.
\begin{definition}[Conditional Expected Marginal Utility of an Item]
\label{def:1}
For any partial realization $\psi$ and any item $e\in E$, the conditional expected marginal utility $\Delta(e \mid \psi)$ of $e$ conditioned on $\psi$ is
\[\Delta(e \mid \psi)=\mathbb{E}_{\Phi}[f(\mathrm{dom}(\psi)\cup \{e\}, \Phi)-f(\mathrm{dom}(\psi), \Phi)\mid \psi \prec \Phi]\]
where the expectation is taken over $\Phi$ with respect to $p(\phi\mid \psi)=\Pr(\Phi=\phi \mid \psi \prec \Phi )$.
\end{definition}


We next introduce the concept of \emph{adaptive submodularity}.
\begin{definition}\cite{golovin2011adaptive}[Adaptive Submodularity]
\label{def:11}
A function $ f: 2^E\times O^E$ is  adaptive submodular with respect to a prior distribution $ p(\phi)$, if for any two partial realizations $\psi$ and $\psi'$ such that $\psi\subseteq \psi'$, and any item $e\in E$ such that $e\notin \mathrm{dom}(\psi')$, the following holds:
\[\Delta(e\mid \psi) \geq \Delta(e\mid \psi') \]
\end{definition}


For comparison purpose, we further introduce the \emph{pointwise submodularity}.
\begin{definition}\cite{golovin2011adaptive}[Pointwise Submodularity]
\label{def:121}
A function $f: 2^E\times O^E\rightarrow \mathbb{R}_{\geq 0}$  is  pointwise submodular if $f(S, \phi)$ is submodular in terms of $S\subseteq E$ for all $\phi\in U^+$. That is, for any $\phi\in U$, any two sets $E_1\subseteq E$ and $E_2 \subseteq E$ such that $E_1 \subseteq E_2$, and any item $e\notin E_2$, we have $f(E_1\cup\{e\}, \phi)- f(E_1, \phi) \geq f(E_2\cup\{e\}, \phi)- f(E_2, \phi)$.
\end{definition}

The above property is referred to as state-wise submodularity in \cite{amanatidis2020fast}. Note that adaptive submodularity does not  imply pointwise submodularity and vice versa.

\begin{algorithm}[t]
\caption{Sampling-based Adaptive Density-Greedy Policy $\pi^{sad}$}
\label{alg:LPP1}
\begin{algorithmic}[1]
\STATE $S_1=\emptyset$, $S_2=\emptyset$, $e^*=\arg\max_{e\in E} f(e)$, $t=1$, $\psi_0=\emptyset$, $C = b$.
\STATE Sample a number $r_0$ uniformly at random from $[0,1]$
\FOR {$e\in E$}
\STATE  let $r_e \sim \mathrm{Bernoulli}(\delta_0)$
\IF {$r_e=1$}
\STATE $S_1=S_1\cup\{e\}$
\ELSE
\STATE $S_2=S_2\cup\{e\}$
\ENDIF
\ENDFOR
\IF [Adopting the first candidate policy $\pi^1$]{$r_0\in[0,\delta_1)$}
\STATE pick $e^*$
\ELSIF [Adopting the second candidate policy $\pi^2$]{$r_0\in[\delta_1, \delta_1 + \delta_2)$}

\STATE $F_1=\{e|e\in S_1, f(e)>0\}$
\WHILE {$F_1 \neq \emptyset $}
\STATE $e_t\leftarrow \arg\max_{e \in F_1}\frac{\Delta(e\mid  \psi_{t-1})}{c_e}$
\STATE select $e_t$ and observe $\Phi(e_t)$
\STATE $\psi_t= \psi_{t-1}\cup\{(e_t, \Phi(e_t))\}$
\STATE $C=C-c_{e_t}$
\STATE $S_1=S_1\setminus\{e_t\}$, $F_1=\{e\in S_1| C\geq c_e \mbox{ and } \Delta(e\mid  \psi_{t-1})>0\}$, $t\leftarrow t+1$
\ENDWHILE
\ELSE[Adopting the third candidate policy $\pi^3$]
\STATE $F_2=\{e|e\in S_2, f(e)>0\}$
\WHILE {$F_2 \neq \emptyset $}
\STATE $e_t\leftarrow \arg\max_{e \in F_2}\frac{\Delta(e\mid  \psi_{t-1})}{c_e}$
\STATE select $e_t$ and observe $\Phi(e_t)$
\STATE $\psi_t= \psi_{t-1}\cup\{(e_t, \Phi(e_t))\}$
\STATE $C=C-c_{e_t}$
\STATE $S_2=S_2\setminus\{e_t\}$, $F_2=\{e\in S_2| C\geq c_e \mbox{ and } \Delta(e\mid  \psi_{t-1})>0\}$, $t\leftarrow t+1$
\ENDWHILE
\ENDIF
\end{algorithmic}
\end{algorithm}
\section{Knapsack Constraint}
\subsection{Algorithm Design}
We first present the design of our \emph{Sampling-based Adaptive Density-Greedy Policy} $\pi^{sad}$ subject to a knapsack constraint.  A detailed description of $\pi^{sad}$ is listed in Algorithm \ref{alg:LPP1}. Our policy is composed of three candidate policies: $\pi^1$, $\pi^2$, and $\pi^3$. The first candidate policy $\pi^1$ selects a singleton  with the maximum expected utility. The other two candidates $\pi^2$ and $\pi^3$ follow a simple density-greedy rule to select items from two random sets respectively. 
Our final policy $\pi^{sad}$ randomly picks one solution from the above three candidates such that $\pi^1$ is selected with probability $\delta_1$, $\pi^2$ is selected with probability  $\delta_2$, and $\pi^3$ is selected with probability $1-\delta_1-\delta_2$. All parameters will be decided later. Although the framework of $\pi^{sad}$  is similar to the modified density greedy algorithm \cite{amanatidis2020fast}, where they only maintain two candidate policies (one is to choose a high value item and the other one is to choose items in a greedy manner), its performance analysis is very
different, as their results hold only if the utility function is both adaptive submodular and pointwise submodular. Later we show that $\pi^{sad}$ achieves a constant approximation ratio without resorting to the property of pointwise submodularity. We next describe $\pi^1$, $\pi^2$, and $\pi^3$ in details.

\textbf{Design of $\pi^1$.} Selecting a singleton  $e^*$ with the maximum expected utility, i.e.,  $e^*=\arg\max_{e\in E} f(e)$.

\textbf{Design of $\pi^2$.} Partition $E$ into two disjoint subsets $S_1$ and $S_2$ (or $E\setminus S_1$) such that $S_1$ contains each item independently with probability $\delta_0$. 
$\pi^2$ selects items only from $S_1$ in a density-greedy manner as follows:
In each round $t$, $\pi^2$ selects an item $e_t$ with the largest ``benefit-to-cost'' ratio from $F_1$ conditioned on the current observation $\psi_{t-1}$
\[e_t\leftarrow \arg\max_{e \in F_1}\frac{\Delta(e\mid  \psi_{t-1})}{c_e}\] where  $F_1=\{e\in S_1| C \geq c_e \mbox{ and } \Delta(e\mid  \psi_{t-1})>0\}$. Here we use $C$ to denote the remaining budget before entering round $t$. After observing the state $\Phi(e_t)$ of $e_t$, we update the partial realization using
$\psi_t= \psi_{t-1}\cup\{(e_t, \Phi(e_t))\}$. This process iterates until $F_1$ becomes an empty set.

\textbf{Design of $\pi^3$.}   Partition $E$ into two disjoint subsets $S_1$ and $S_2$ (or $E\setminus S_1$) such that $S_1$ contains each item independently with probability $\delta_0$. 
$\pi^3$ selects items only from $S_1$ in the same density-greedy manner as used in the design of $\pi^2$.

\subsection{Performance Analysis}
 Note that all existing results  on non-monotone adaptive submodular maximization \cite{amanatidis2020fast,hanicml} require Lemma 4 of \cite{gotovos2015non}, whose proof relied on the assumption of the pointwise submodularity of the utility function. To relax this assumption, we first provide two technical lemmas, whose proofs do not require pointwise submodularity. We use $\mathrm{range}(\pi)$ to denote the set containing all items that $\pi$ selects for some $\phi\in U^+$, i.e., $\mathrm{range}(\pi)=\{e | e\in \cup_{\phi\in U^+} E(\pi, \phi)\}$.
\begin{lemma}
\label{lem:2}
If $f: 2^E\times O^E\rightarrow \mathbb{R}_{\geq 0}$ is adaptive submodular with respect to $p(\phi)$, then for any three policies $\pi^{a}$, $\pi^{b}$, and $\pi^{c}$ such that $\mathrm{range}(\pi^{b})\cap \mathrm{range}(\pi^{c})=\emptyset$, we have
\begin{eqnarray}
f_{avg}(\pi^a@ \pi^b) + f_{avg}(\pi^a @\pi^c)\geq  f_{avg}(\pi^a)
\end{eqnarray}
\end{lemma}
\emph{Proof:} For each $r\in\{a, b, c\}$, let $\overrightarrow{\psi^r}=\{\psi^{r}_0, \psi^{r}_1, \psi^{r}_2,\cdots, \psi^{r}_{|\overrightarrow{\psi^r}|-1}\}$ denote a fixed run of $\pi^r$, where for each $t\in [|\overrightarrow{\psi^r}|-1]$, $\psi^{r}_t$ is the partial realization of the first $t$ selected items. For ease of presentation, let $\psi^{r}$ denote the final observation $\psi^{r}_{|\overrightarrow{\psi^r}|-1}$ of $\overrightarrow{\psi^r}$  for short. 
 For each $e\in \mathrm{range}(\pi^{c})$ and $t\in[n-1]$, let $I(\pi^c, e, t+1)$ be indicator variable  that $e$ is selected as the $(t+1)$-th item by $\pi^c$. Let $\overrightarrow{\Psi^a}, \overrightarrow{\Psi^b}, \overrightarrow{\Psi^c}$ denote random realizations of $\overrightarrow{\psi^a}, \overrightarrow{\psi^b}, \overrightarrow{\psi^c}$, respectively. Then we have
{\small \begin{eqnarray}
&&f_{avg}(\pi^a@ \pi^c)- f_{avg}(\pi^a)~\nonumber \\
&&=\mathbb{E}_{(\overrightarrow{\Psi^a}, \overrightarrow{\Psi^c})}[
\sum_{e\in \mathrm{range}(\pi^{c}), t\in [|\overrightarrow{\Psi^c}|-2]}\mathbb{E}[I(\pi^c, e, t+1)|\Psi^{c}_t]\Delta(e\mid\Psi^{a}\cup \Psi^{c}_t)]~\nonumber\\
&&=\mathbb{E}_{(\overrightarrow{\Psi^a}, \overrightarrow{\Psi^b}, \overrightarrow{\Psi^c})}[
\sum_{e\in \mathrm{range}(\pi^{c}), t\in [|\overrightarrow{\Psi^c}|-2]}\mathbb{E}[I(\pi^c, e, t+1)|\Psi^{c}_t]\Delta(e\mid\Psi^{a}\cup \Psi^{c}_t)]\label{eq:ui}
\end{eqnarray}
where the first expectation is taken over the prior joint distribution of $\overrightarrow{\psi^a}$ and $\overrightarrow{\psi^c}$ and the second expectation is taken over the prior joint distribution of $\overrightarrow{\psi^a}$, $\overrightarrow{\psi^b}$and $\overrightarrow{\psi^c}$.
  \begin{eqnarray}
&&f_{avg}(\pi^a@ \pi^b @\pi^c)- f_{avg}(\pi^a@\pi^b)~\nonumber \\
&&=\mathbb{E}_{(\overrightarrow{\Psi^a}, \overrightarrow{\Psi^b}, \overrightarrow{\Psi^c})}[
\sum_{e\in \mathrm{range}(\pi^{c}), t\in [|\overrightarrow{\Psi^c}|-2]}\mathbb{E}[I(\pi^c, e, t+1)|\Psi^{c}_t]\Delta(e\mid \Psi^{a}\cup \Psi^{b}\cup \Psi^c_t)]\label{eq:uii}
\end{eqnarray}}
where the expectation is taken over the prior joint distribution of $\overrightarrow{\psi^a}$, $\overrightarrow{\psi^b}$and $\overrightarrow{\psi^c}$.

Because $f: 2^E\times O^E\rightarrow \mathbb{R}_{\geq 0}$ is adaptive submodular with respect to $p(\phi)$, and $\mathrm{range}(\pi^{b})\cap \mathrm{range}(\pi^{c})=\emptyset$, then for any given realizations $(\overrightarrow{\psi^a}, \overrightarrow{\psi^b}, \overrightarrow{\psi^c})$ after running $\pi^a@ \pi^b @\pi^c$, any $t\in [|\overrightarrow{\psi^c}|-2]$, and any item $e\in \mathrm{range}(\pi^{c})$, we have $\Delta(e\mid \psi^{a}\cup \psi^{c}_t) \geq \Delta(e\mid \psi^{a}\cup \psi^{b}\cup \psi^c_t)$ due to $e\notin \mathrm{dom}(\psi^{b})$. This together with (\ref{eq:ui}) and (\ref{eq:uii}) implies that
\begin{eqnarray*}
f_{avg}(\pi^a@ \pi^c)- f_{avg}(\pi^a) \geq f_{avg}(\pi^a@ \pi^b @\pi^c)- f_{avg}(\pi^a@\pi^b)
\end{eqnarray*}

Hence,
\begin{eqnarray}
&&f_{avg}(\pi^a@ \pi^b @\pi^c)~\nonumber\\
&=&f_{avg}(\pi^a)+(f_{avg}(\pi^a@ \pi^b)- f_{avg}(\pi^a)) + (f_{avg}(\pi^a@ \pi^b @\pi^c)- f_{avg}(\pi^a@\pi^b))~\nonumber\\
&\leq& f_{avg}(\pi^a)+(f_{avg}(\pi^a@ \pi^b)- f_{avg}(\pi^a)) + (f_{avg}(\pi^a@ \pi^c)- f_{avg}(\pi^a))\label{eq:4}
\end{eqnarray}
Because $f_{avg}(\pi^a@ \pi^b @\pi^c)\geq0$, we have
\begin{eqnarray}
f_{avg}(\pi^a)+(f_{avg}(\pi^a@ \pi^b)- f_{avg}(\pi^a)) + (f_{avg}(\pi^a@ \pi^c)- f_{avg}(\pi^a))\geq0 \label{eq:5}
\end{eqnarray}
It follows that
{\small \begin{eqnarray*}
&&f_{avg}(\pi^a@ \pi^b) + f_{avg}(\pi^a @\pi^c)\\
&&=f_{avg}(\pi^a)+(f_{avg}(\pi^a@ \pi^b)- f_{avg}(\pi^a)) + f_{avg}(\pi^a) + (f_{avg}(\pi^a@ \pi^c)- f_{avg}(\pi^a))\\
&&\geq  f_{avg}(\pi^a)
\end{eqnarray*}}
The inequality is due to (\ref{eq:5}). $\Box$

We next present the second technical lemma.
\begin{lemma}
\label{lem:1}
Let $\pi\in\Omega$ denote a policy that selects items from $S$ in the same density-greedy manner as used in the design of $\pi^2$ and $\pi^3$, where $S$ is a random set that is obtained by independently picking each item with probability $\sigma$. If $f: 2^{E}\times O^E\rightarrow \mathbb{R}_{\geq0}$ is adaptive submodular with respect to $p(\phi)$, then
\begin{eqnarray}
\label{eq:2}
(2+\frac{1}{\sigma})f_{avg}(\pi) + f(e^*) \geq  f_{avg}(\pi^{opt}@\pi)
\end{eqnarray}
\end{lemma}
\emph{Proof:} In the proof of Theorem 4 of \cite{amanatidis2020fast}, they show that (\ref{eq:2}) holds
 if $f: 2^{E}\times O^E\rightarrow \mathbb{R}_{\geq0}$  is adaptive submodular and pointwise submodular with respect to $p(\phi)$. In fact, we can prove a more general result by relaxing the assumption about the property of pointwise submodularity, i.e., we next show that (\ref{eq:2}) holds  if $f: 2^{E}\times O^E\rightarrow \mathbb{R}_{\geq0}$ is adaptive submodular with respect to $p(\phi)$. By inspecting the proof of  (\ref{eq:2}) of \cite{amanatidis2020fast}, it is easy to find that the only part that requires the property of pointwise submodularity is the proof of Lemma 5. We next show that Lemma 5 holds without resorting to pointwise submodularity.

Before restating Lemma 5 in \cite{amanatidis2020fast}, we introduce some notations. Let $\overrightarrow{\psi}=\{\psi_0, \psi_1, \psi_2,\cdots, \psi_{|\overrightarrow{\psi}|-1}\}$ denote a fixed run of $\pi$, where for any $t\in[|\overrightarrow{\psi}|-1]$, $\psi_t$ represents the partial realization of the first $t$ selected items. For notation simplicity, for every $\overrightarrow{\psi}$, let $\psi'$ denote  the final observation $\psi_{|\overrightarrow{\psi}|-1}$ for short. Note that in \cite{amanatidis2020fast}, they defer the ``sampling'' phase of $\pi$ without affecting the distributions of the output, that is, they toss a coin of success $\sigma$ to decide whether or not to add an item to the solution  each time after an item is being considered.  Let $M(\overrightarrow{\psi})$ denote those items which are considered but not chosen by $\pi$ which have positive expected marginal contribution
to $\psi'$ under $\overrightarrow{\psi}$. Lemma 5 in \cite{amanatidis2020fast} states that $f_{avg}(\pi)  \geq \sigma\times  \sum_{\overrightarrow{\psi}}\Pr[\overrightarrow{\psi}] \sum_{e\in M(\overrightarrow{\psi})}\Delta( e \mid \psi')$ where $\Pr[\overrightarrow{\psi}]$ is the probability that $\overrightarrow{\psi}$ occurs.

For any item $e\in E$, let $\Lambda_e$ denote the set of all possible partial realizations $\psi$ such that $\psi$ is observed right before $e$ is being considered, i.e., $\Lambda_e = \{\psi\mid \mbox{$\psi$ is the last observation before $e$ is being considered}\}$. Let $\mathcal{D}_e$ denote the prior distribution over all partial realizations in $\Lambda_e$, i.e., for each $\psi\in \Lambda_e$, $\mathcal{D}_e(\psi)$ is the probability that $e$ has been considered and $\psi$ is the last observation before  $e$ is being considered. It follows that
\begin{eqnarray*}
f_{avg}(\pi)  &=& \sum_{e\in E} \mathbb{E}_{\Psi\sim \mathcal{D}_e}[\sigma\times \Delta( e \mid \Psi)]\\
&\geq& \sum_{e\in E} \mathbb{E}_{\Psi\sim \mathcal{D}_e}[\sigma\times \sum_{\overrightarrow{\psi}}\Pr[\overrightarrow{\psi}\mid \Psi, e]  \Delta( e \mid \psi')]\\
&=& \sigma\times \sum_{e\in E} \mathbb{E}_{\Psi\sim \mathcal{D}_e}[ \sum_{\overrightarrow{\psi}}\Pr[\overrightarrow{\psi}\mid \Psi, e]  \Delta( e \mid \psi')]\\
&=& \sigma\times  \sum_{\overrightarrow{\psi}}\Pr[\overrightarrow{\psi}] (\sum_{e\in M(\overrightarrow{\psi})}\Delta( e \mid \psi'))
\end{eqnarray*}
where $\Pr[\overrightarrow{\psi}\mid \Psi, e]$ is the probability that $\overrightarrow{\psi}$ occurs conditioned on the event that $e$ is being considered and  $\Psi$ is the last observation before $e$ is being considered. The first equality is due to the assumption that each item is sampled with probability $\sigma$, and the  inequality is due to the observations that $e\notin \mathrm{dom}(\psi')$, $\Psi \subseteq \psi'$ and $f: 2^{E}\times O^E\rightarrow \mathbb{R}_{\geq0}$  is adaptive submodular. $\Box$

Lemma \ref{lem:1}, together with the design of $\pi^2$ and $\pi^3$, implies the following two corollaries.
\begin{corollary}\label{cor:1}
If $f: 2^{E}\times O^E\rightarrow \mathbb{R}_{\geq0}$ is adaptive submodular with respect to $p(\phi)$, then $(2+\frac{1}{\delta_0})f_{avg}(\pi^2) + f(e^*) \geq  f_{avg}(\pi^{opt}@\pi^2)$.
\end{corollary}
\begin{corollary}\label{cor:2}
If $f: 2^{E}\times O^E\rightarrow \mathbb{R}_{\geq0}$ is adaptive submodular with respect to $p(\phi)$, then $(2+\frac{1}{1-\delta_0})f_{avg}(\pi^3) + f(e^*) \geq  f_{avg}(\pi^{opt}@\pi^3)$.
\end{corollary}

Now we are ready to present the main theorem of this section.
\begin{theorem}If $f: 2^{E}\times O^E\rightarrow \mathbb{R}_{\geq0}$ is adaptive submodular with respect to $p(\phi)$, then for $\delta_1=1/5$, $\delta_2=2/5$, and $\delta_0=1/2$, we have $f_{avg}(\pi^{sad})\geq  \frac{1}{10}f_{avg}(\pi^{opt})$.
\end{theorem}
\emph{Proof:} Recall that $\pi^2$ and $\pi^3$ start with a random partition of $E$ into two disjoint subsets according to the same distribution. It is safe to assume that $\pi^2$ and $\pi^3$ share a common phase of generating such a partition as this assumption does not affect the expected utility of either $\pi^2$ or $\pi^3$. Thus, given a fixed partition $(S_1, S_2)$, $\pi^2$ and $\pi^3$ are running on two disjoint subsets because $\pi^2$ selects items only from $S_1$ and $\pi^3$ selects items only from $S_2$.  It follows that $\mathrm{range}(\pi^2)\cap \mathrm{range}(\pi^3)=\emptyset$ conditional on any fixed pair of $(S_1, S_2)$. Letting $\mathbb{E}[f_{avg}(\pi^{opt}@\pi^2)+f_{avg}(\pi^{opt}@\pi^3)| (S_1, S_2)]$ denote the conditional expected value of $f_{avg}(\pi^{opt}@\pi^2)+f_{avg}(\pi^{opt}@\pi^3)$ conditioned on $(S_1, S_2)$, Lemma \ref{lem:2} implies that for any fixed pair of  $(S_1, S_2)$,
\begin{eqnarray}
&&\mathbb{E}[f_{avg}(\pi^{opt}@\pi^2)+f_{avg}(\pi^{opt}@\pi^3)| (S_1, S_2)] \geq \mathbb{E}[f_{avg}(\pi^{opt})| (S_1, S_2)]~\nonumber\\
&&= f_{avg}(\pi^{opt})\label{eq:89}
\end{eqnarray}
The equality is due to the observation that the expected utility of the optimal solution $\pi^{opt}$ is independent of the realizations of $S_1$ and $S_2$. Taking the expectation of $\mathbb{E}[f_{avg}(\pi^{opt}@\pi^2)+f_{avg}(\pi^{opt}@\pi^3)| (S_1, S_2)]$ over $(S_1, S_2)$, (\ref{eq:89}) implies that
\begin{eqnarray}
&&f_{avg}(\pi^{opt}@\pi^2)+f_{avg}(\pi^{opt}@\pi^3) \geq f_{avg}(\pi^{opt}) \label{eq:6}
\end{eqnarray}
Hence,
\begin{eqnarray}
&&(2+\frac{1}{\delta_0})f_{avg}(\pi^2) + f(e^*) + (2+\frac{1}{1-\delta_0})f_{avg}(\pi^3) + f(e^*) ~\nonumber\\
&&\geq  f_{avg}(\pi^{opt}@\pi^2)+f_{avg}(\pi^{opt}@\pi^3)~\nonumber\\
&&\geq f_{avg}(\pi^{opt}) \label{eq:1}
\end{eqnarray}
The first inequality is due to Corollary \ref{cor:1}  and  Corollary \ref{cor:2}. The second inequality is due to (\ref{eq:6}). Because $f(e^*) = f_{avg}(\pi^1)$, (\ref{eq:1}) implies that
\begin{eqnarray*}
(2+\frac{1}{\delta_0})f_{avg}(\pi^2) + (2+\frac{1}{1-\delta_0})f_{avg}(\pi^3) + 2f_{avg}(\pi^1) \geq f_{avg}(\pi^{opt})
\end{eqnarray*}

Recall that $\pi^{sad}$ randomly picks one solution from  $\{\pi^1, \pi^2, \pi^3\}$ such that $\pi^1$ is picked with probability $\delta_1$, $\pi^2$ is picked with probability  $\delta_2$, and $\pi^3$ is picked with probability $1-\delta_1-\delta_2$.  If we set $\delta_1 = \frac{2}{(2+\frac{1}{\delta_0}) + (2+\frac{1}{1-\delta_0}) + 2}$ and $\delta_2 = \frac{2+\frac{1}{\delta_0}}{(2+\frac{1}{\delta_0}) + (2+\frac{1}{1-\delta_0}) + 2}$, then we have
{\small\begin{eqnarray*}
 f_{avg}(\pi^{sad}) = &&\frac{2+\frac{1}{\delta_0}}{(2+\frac{1}{\delta_0}) + (2+\frac{1}{1-\delta_0}) + 2}f_{avg}(\pi^2) + \frac{2+\frac{1}{1-\delta_0}}{(2+\frac{1}{\delta_0}) + (2+\frac{1}{1-\delta_0}) + 2}f_{avg}(\pi^3) \\
 &&+ \frac{2}{(2+\frac{1}{\delta_0}) + (2+\frac{1}{1-\delta_0}) + 2}f_{avg}(\pi^1) \geq \frac{1}{6+\frac{1}{\delta_0}+\frac{1}{1-\delta_0}} f_{avg}(\pi^{opt})
\end{eqnarray*}}
If we set $\delta_0=1/2$, then $f_{avg}(\pi^{sad})\geq \frac{1}{10}f_{avg}(\pi^{opt})$. $\Box$

\emph{Remark:} Recall that under the optimal setting, $\delta_0=1/2$, which indicates that $\pi^2$ is identical to $\pi^3$. Thus, we can simplify the design of $\pi^{sad}$ to maintain only two candidate policies $\pi^1$ and $\pi^2$. In particular, given that $\delta_1=1/5$ and $\delta_2=2/5$ under the optimal setting, $\pi^{sad}$ randomly picks a policy from $\pi^1$ and $\pi^2$ such that  $\pi^1$ is picked with probability $1/5$ and $\pi^2$ is picked with probability $4/5$. It is easy to verify that this simplified version of $\pi^{sad}$ and its original version have identical output distributions.

\begin{algorithm}[t]
\caption{Sampling-based Adaptive Greedy Policy $\pi^{sag}$}
\label{alg:LPP2}
\begin{algorithmic}[1]
\STATE $S_1=\emptyset$, $S_2=\emptyset$, $\psi_0=\emptyset$, $V=\emptyset$.
\STATE Sample a number $r_0$ uniformly at random from $[0,1]$
\FOR {$e\in E$}
\STATE  let $r_e \sim \mathrm{Bernoulli}(\delta_0)$
\IF {$r_e=1$}
\STATE $S_1=S_1\cup\{e\}$
\ELSE
\STATE $S_2=S_2\cup\{e\}$
\ENDIF
\ENDFOR
\IF [Adopting the first candidate policy $\pi^1$]{$r_0\in[0,\delta_1)$}
\STATE $F_1=\{e|e\in S_1, f(e)>0\}$
\WHILE {$F_1 \neq \emptyset $}
\STATE $e_t\leftarrow \arg\max_{e \in F_1}\Delta(e\mid  \psi_{t-1})$
\STATE select $e_t$ and observe $\Phi(e_t)$
\STATE $\psi_t= \psi_{t-1}\cup\{(e_t, \Phi(e_t))\}$
\STATE $V\leftarrow V\cup\{e_t\}$
\STATE $S_1=S_1\setminus\{e_t\}$, $F_1=\{e\in S_1| V\cup\{e\}\in \mathcal{I} \mbox{ and } \Delta(e\mid  \psi_{t-1})>0\}$, $t\leftarrow t+1$
\ENDWHILE
\ELSE[Adopting the second candidate policy $\pi^2$]
\STATE $F_2=\{e|e\in S_2, f(e)>0\}$
\WHILE {$F_2 \neq \emptyset $}
\STATE $e_t\leftarrow \arg\max_{e \in F_2}\Delta(e\mid  \psi_{t-1})$
\STATE select $e_t$ and observe $\Phi(e_t)$
\STATE $\psi_t= \psi_{t-1}\cup\{(e_t, \Phi(e_t))\}$
\STATE $V\leftarrow V\cup\{e_t\}$
\STATE $S_2=S_2\setminus\{e_t\}$, $F_2=\{e\in S_2| V\cup\{e\}\in \mathcal{I} \mbox{ and } \Delta(e\mid  \psi_{t-1})>0\}$, $t\leftarrow t+1$
\ENDWHILE
\ENDIF
\end{algorithmic}
\end{algorithm}
\section{$k$-System Constraint}
\subsection{Algorithm Design}
We next present a \emph{Sampling-based Adaptive Greedy Policy} $\pi^{sag}$ subject to a $k$-system constraint.  A detailed description of $\pi^{sag}$ is listed in Algorithm \ref{alg:LPP2}. 
 $\pi^{sag}$ randomly picks one solution from two candidate policies, $\pi^1$ and $\pi^2$, such that $\pi^1$ is selected with probability $\delta_1$, $\pi^2$ is selected with probability  $1-\delta_1$.  Both  $\pi^1$ and $\pi^2$ follow a simple greedy rule to select items from two random sets respectively. We next describe $\pi^1$ and $\pi^2$ in details. All parameters will be optimized later.


\textbf{Design of $\pi^1$.}  Partition $E$ into two disjoint subsets $S_1$ and $S_2$ (or $E\setminus S_1$) such that $S_1$ contains each item independently with probability $\delta_0$. $\pi^1$ selects items only from $S_1$ in a greedy manner as follows:
In each round $t$, $\pi^1$ selects an item $e_t$ with the largest marginal value from $F_1$ conditioned on the current observation $\psi_{t-1}$
\[e_t\leftarrow \arg\max_{e \in F_1}\Delta(e\mid  \psi_{t-1})\] where  $F_1=\{e\in S_1| V\cup\{e\}\in \mathcal{I} \mbox{ and } \Delta(e\mid  \psi_{t-1})>0\}$. Here $V$ denotes the first $t-1$ items selected by $\pi^1$. After observing the state $\Phi(e_t)$ of $e_t$, we update the partial realization using
$\psi_t= \psi_{t-1}\cup\{(e_t, \Phi(e_t))\}$.  This process iterates until $F_1$ becomes an empty set.

\textbf{Design of $\pi^2$.} Partition $E$ into two disjoint subsets $S_1$ and $S_2$ (or $E\setminus S_1$) such that $S_1$ contains each item independently with probability $\delta_0$. $\pi^2$ selects  items only from $S_2$ in the same greedy manner as used in the design of $\pi^1$. 

\subsection{Performance Analysis}
Before presenting the main theorem, we first provide a technical lemma from \cite{hanicml}.
\begin{lemma}\cite{hanicml}
\label{lem:111}
Let $\pi\in\Omega$ denote a feasible $k$-system constrained policy that  chooses items from $S$  in the same greedy manner as used in the design of $\pi^1$ and $\pi^2$, where $S$ is a random set that is obtained by independently picking each item with probability $\sigma$. If $f: 2^{E}\times O^E\rightarrow \mathbb{R}_{\geq0}$ is adaptive submodular with respect to $p(\phi)$, then
$
(k+\frac{1}{\sigma})f_{avg}(\pi) \geq  f_{avg}(\pi^{opt}@\pi)$.
\end{lemma}

Lemma \ref{lem:111}, together with the design of $\pi^1$ and $\pi^2$, implies the following two corollaries.
\begin{corollary}\label{cor:111}
If $f: 2^{E}\times O^E\rightarrow \mathbb{R}_{\geq0}$ is adaptive submodular with respect to $p(\phi)$, then
$
(k+\frac{1}{\delta_0})f_{avg}(\pi^1) \geq  f_{avg}(\pi^{opt}@\pi^1)$.
\end{corollary}
\begin{corollary}\label{cor:222}
If $f: 2^{E}\times O^E\rightarrow \mathbb{R}_{\geq0}$ is adaptive submodular with respect to $p(\phi)$, then
$(k+\frac{1}{1-\delta_0})f_{avg}(\pi^2) \geq  f_{avg}(\pi^{opt}@\pi^2)$.
\end{corollary}

Now we are ready to present the main theorem of this section.
\begin{theorem}If $f: 2^{E}\times O^E\rightarrow \mathbb{R}_{\geq0}$ is adaptive submodular with respect to $p(\phi)$, then for $\delta_1=1/2$ and $\delta_0=1/2$, we have
$
f_{avg}(\pi^{sag})\geq  \frac{1}{2k+4}f_{avg}(\pi^{opt})$.
\end{theorem}
\emph{Proof:} Following the same proof as for (\ref{eq:6}),  we have
\begin{eqnarray}
&&f_{avg}(\pi^{opt}@\pi^1)+f_{avg}(\pi^{opt}@\pi^2) \geq f_{avg}(\pi^{opt}) \label{eq:66}
\end{eqnarray}
Hence,
\begin{eqnarray}
&&(k+\frac{1}{\delta_0})f_{avg}(\pi^1) + (k+\frac{1}{1-\delta_0})f_{avg}(\pi^2)~\nonumber\\
&&\geq  f_{avg}(\pi^{opt}@\pi^1)+f_{avg}(\pi^{opt}@\pi^2)~\nonumber\\
&&\geq f_{avg}(\pi^{opt}) \label{eq:111}
\end{eqnarray}
The first inequality is due to Corollary \ref{cor:111}  and  Corollary \ref{cor:222}. The second inequality is due to (\ref{eq:66}).

Recall that $\pi^{sag}$ randomly picks one solution from  $\{\pi^1, \pi^2\}$ such that $\pi^1$ is picked with probability $\delta_1$ and $\pi^2$ is picked with probability  $1-\delta_1$.  If we set $\delta_1 = \frac{k+\frac{1}{\delta_0}}{(k+\frac{1}{\delta_0}) + (k+\frac{1}{1-\delta_0})}$, then
\begin{eqnarray*}
 f_{avg}(\pi^{sag}) = &&\frac{k+\frac{1}{\delta_0}}{(k+\frac{1}{\delta_0}) + (k+\frac{1}{1-\delta_0})}f_{avg}(\pi^1) + \frac{k+\frac{1}{1-\delta_0}}{(k+\frac{1}{\delta_0}) + (k+\frac{1}{1-\delta_0})}f_{avg}(\pi^2) \\
 &&\geq \frac{1}{(k+\frac{1}{\delta_0}) + (k+\frac{1}{1-\delta_0})} f_{avg}(\pi^{opt})
\end{eqnarray*}
If we set $\delta_0=1/2$, then $f_{avg}(\pi^{sag})\geq \frac{1}{2k+4}f_{avg}(\pi^{opt})$. $\Box$

\emph{Remark:} Recall that under the optimal setting, $\delta_0=1/2$, which indicates that $\pi^1$ is identical to $\pi^2$. Thus, we can simplify the design of $\pi^{sag}$ such that it maintains only one policy $\pi^1$. It is easy to verify that this simplified version of $\pi^{sag}$ and its original version have identical output distributions.


\bibliographystyle{ijocv081}
\bibliography{reference}




\end{document}